\documentclass[12pt, letterpaper]{article}
\pdfoutput=1
\usepackage{amsmath}
\usepackage{graphicx}
\usepackage{natbib}
\usepackage[margin=1.0in]{geometry}
\usepackage{setspace}
\providecommand{\keywords}[1]{\textbf{\textit{Keywords:}} #1}
\usepackage{authblk}

\title{How polarization can provide an increase in content dissemination amongst the highly ranked influencers}

\author{Cameron E. Taylor$^1$ \and Ivan Garibay$^2$ \and Alexander V. Mantzaris$^3$}

\date{
  $^1$ Research Assistant (UCF), cetaylor758903@knights.ucf.edu, 239-628-7890, University of Central Florida , 4000 Central Florida Blvd
P.O. Box 162370, Orlando, FL 32816-2370\\
  $^2$ Assistant Professor / Lab Director (UCF), igaribay@ucf.edu, 407-823-2204, Engr. II Room 424  University of Central Florida, 4000 Central Florida Blvd
P.O. Box 162370, Orlando, FL 32816-2370 \\
  $^3$ Assistant Professor (UCF), alexander.mantzaris@ucf.edu, 407-823-3631;  TCII 211A University of Central Florida, 4000 Central Florida Blvd
P.O. Box 162370, Orlando, FL 32816-2370 \\
}
\begin{document}
\maketitle

 \newpage

\begin{abstract}
  This work extends a model of simulating influence in a network of stochastic edge dynamics to account for polarization. The model built upon  is termed \emph{Dynamic Communicators} and seeks to understand the process which produces low volume high influence amongst users. This model is extended to introduce the effects polarization. The fundamental assumption of the model is that a parameter of \emph{importance} governs the rate of message responsiveness. With the introduction of relative incremental changes according to the response incurred in adjacent nodes receiving content, the changes in the power brokerage of a network can be examined. This provides a content agnostic interpretation for the desire to proliferate content amongst peers. From the results of the simulations, the analysis shows that a lack of polarization incrementally develops a more level discussion network with more even response rates whereas the polarization introduction leads to a gradual increase in response rate disparity.

\end{abstract}

\newpage

\section{Introduction}
There has been an increasing interest in the analysis of polarization in social networks \cite{sinclair2014party,stonecash2018diverging} as a larger proportion of the political discourse happens online. Indicators of instability in social networks can be seen as forward predictors of potential disruptions in the political decision making amongst citizens \cite{abramowitz2015new} (shows that the disagreement in political ideologies is increasing over time in the USA). The work of \cite{baldassarri2007dynamics,baldassarri2008partisans} supports the observed trend, and motivates the study of the topic from a modeling perspective to infer the fundamental dynamics of the process in the hope that hypothetical negative consequences can be avoided. Polarization can lead to the break of communities as described in the historic example of the Zachary karate club network \cite{zachary1977information} (a group of a sports club through a disagreement created a split/separation). The Zachary karate club example shows how the monitoring of the network structure and dynamics can be used to predict a break which on a larger scale could signal a much larger set of consequences. To understand the underlying process in more detail we develop a model to capture the effects of the communication influence a network has before and after a polarization event to determine whether certain members have a potential return from the introduction of a disagreement.

A recent model that analyzes the formation of the polarization process is, \cite{davies2017homophily}. It uses the basic concepts of homophily and accessiblity as potential links within a subset pool of potential new contacts; as a means to discover homophilic edges towards users. The potential to complete a local network by pooling from enough users with similar features is taken as a basic premise in the edge creation for the network. A simulation for the change of opinions is modeled with a binary voter model \cite{clifford1973model}. Although this captures the competition for ideas to be reflected upon a node locality it does not fully represent different magnitudes of influence that are accumulated by the 'network effect' over time via the message exchange mechanisms. The desire to control a network appears to be in many ways a possible asset as discussed to a great extent in \cite{benkler2006wealth} (The wealth of networks). Obtaining control of a network is observed to be something actively sought out by members, and succeeding is presumably a transfer of this 'wealth'.

In this paper we look at the effects of an established even split between members of a messaging exchange network of sizes where it is considered feasible that all members can send messages directly to each other. The concern is on whether the mechanisms of a negative view of content originating from members expresssing an opposing view of a discourse will incur a direction for understanding motives and incentives. Given a user is participating in actions which may result in a fractioning of their community would the result bring a question to certain members who must ask whether the result will beneficial to them or not after the continuation of the polarization behaviors. This benefit  is considered  to be related to whether they can be expected to have a greater contribution to their community from the message exchanges they participate in. The results will be analyzed by the extent of the influence they have on their peers to respond to their messages by propagating them. Instead of assuming a state variable change, as in the binary voting model, the message propogation is considered to be an indicator of a state change for the ideological positioning. The discussion leaders can be seen as a source of information for the group and if the responses are focused upon a small or large portion of the members, different overall dynamics can be interpreted. Recent research on the political tensions in Catalonia look into whether this polarization is driven by a small group of people with large societal influence \cite{barrio2017reducing} which can provide insight into the nature of communications of highly leveraged influence on opinions for societal decision making when a distribution is far from uniform. The difference between homophily and influence is examined on a large messaging network in \cite{aral2009distinguishing} and the findings show that influence plays a smaller role than homophily. Therefore, in the equations for the firing rules in this work, the homophily is the driving force for the influence. This is one of the motivations behind this research. The state of a node's affiliation is taken into account during  the message response generation rather than attempting the inference of a latent variable under an specific representation space.

The model of dynamic communicators \cite{mantzaris2012model} replicates the effect of efficient power brokering over a temporal dynamic network defined over a set of transient edges induced from direct messaging. In this model the nodes of the network proceed to send messages via a stochastic message creation and dissemination mechanism. The model represents the dynamics for creating and spreading content as well as the activity decision for a node's engagement when a directed edge is created towards it (receives a message). The factor determining the action that a node has once a  message is received (directed edge towards it) and whether that content is promoted through it subsequently as a proxy for its spread is determined by a parameter of \emph{importance}. A message is likely to be spread if it originates from a node with a large allocated importance relative to the other network members. The message is assumed to be without memory so that although the original content creation can be from the discussion leader, those at the bottom of the importance hierarch cannot utilize this fact for subsequent content disseminations. The modeling perspective takes into account piggy backing where repetition and coinciding discussions within a temporal dependency can be utilized to self promote and is explained in the methodology section.

This direction focuses upon building a model which develops macrodynamics from the 'bottom-up', \cite{katz1955lazarsfeld,schelling2006micromotives} which  is used to uncover a simple model which portrays many of the phenomena we observe in practice. The message passing framework is chosen to encapsulate the social modeling paradigm as a useful abstraction. This is shown in the work of \cite{hedstrom2005dissecting} which emphasizes the mechanisms of entities and activities being organized to bring about the change from their inter-linking.  The disruption in the pathways of information \cite{kossinets2008structure} for the communities to communicate with other members is assumed to be reflected in the probabilities to create content as a result of exposure due to a node with that is aligned with their local view of what is included or excluded and support for the polarity representation used in our model.

This modeling approach of an importance factor governs the firing rate for edge connections and fits the influence dynamics observed in the dataset of ENRON \cite{klimt2004enron} where the influential effects of topp tier employees produced long downstream propagations of content. This observation was measured using the methodologies of \cite{grindrod2011communicability,grindrod2014dynamical}. We use the foundation of the model of dynamic communicators to build upon it to explain the effects of importance value changes over the evolution of the model. A mechanism for incrementing the main factor of influence based on the history of participation in message disseminations introduced with single step memory assumptions. Increments will produce an increased chance for other members of the network to respond to their content via replication and dissemination.

The work in \cite{baldassarri2007dynamics} develops a formal model of social influence for microinteractions which leads to macrostructural outcomes regarding the perception of polarization. It encapsulates the concept that personal influence can be directed from sensitive discussions arising from opinions of diverse issues. Our firing equation for the influence propogation in Dynamic communicators has analogous forms in research such as eq.1 of the paper. The choice for the max of the denominator in relation to the edge in question, is seen there as well for normalization. In this model we do not represent the topics but the affiliation of the polar sides which is common in community disfunction.

Their model describes a stochastic process where influence is based on a distance measure from a set of local actors, which avoids an explicit spatial representation as \cite{axelrod1997dissemination}. The work here also removes the explicit spatial structure for the model interactions as it resembles more of the data analytics gathered. This is largely due to privacy concerns and that most of the data is too sparse in regards to the density required to perform an accurate spatial temporal modeling of the micro interactions. A key aspect of this model and related ones is that the mechanism for preference does not impose connectivity constraints. Our work is also based on the assumption that participants of a polar or non-polarized community can interact equivalently.

Temporal network analysis has many methodologies to be used in order to  measure aspects of the information contained in the datasets, \cite{holme2012temporal}. The ability for messages to be spread around a network and the ability to initiate message responses from peers in a community is assumed to be indicator of influence to be captured from the stochastic simulations produced here as in various methodologies in the area. The applications of these principles applies widely and to a diverse set of datasets such as its effectiveness for analyzing the connectivity of phone calls between academics \cite{eagle2009inferring} where the edges are sparse in comparison to the duration and content density of the presence of the edge.

\section{Methodology}

The work of Dynamic Communicators \cite{mantzaris2012model} offers an intuitively reasonable explanatory mechanism for how the phenomena measured from a complete network (members have full visibility and communication channels of all members) can arise to produce dynamic centrality. Namely the effect where nodes producing low numbers of messages which spread to a large number of nodes indirectly through intermediaries/relays. This is done without explicitly defining a differentiation of the messages that favors a proportion of edges to have this property and therefore allows for a homogeneous event space.
Information exchanges which have temporal dependencies rely upon a series of knock on effects. It is assumed that these events of a response are generated through a local decision criteria and are not aware of the mean field effect of the identical content impact elsewhere in the network. Although the individual considerations regarding a node's discretion on whether to propogate a message it receives or not is a complex process the  macro behavior of the knock on effect is the main focus. The parameterization for these decisions are represented as a parameter  of each node which is its \emph{importance value} ranked among the other nodes; $l_n: 0 < l_1 \leq l_2 \leq \ldots \leq l_N$. Although this may seem simplistic, it does represent a ranked list for characteristics of popularity, influence, and other effects such as authority when this is directly linked to the response of a content receiving event.

It is assumed there is a \emph{Basal} rate for which nodes generate new content, $b$, as a probability which is uniform across the network and time. These events produce $c_b$ events uniformly distributed to all the other nodes. Each event is realized as the production of a directed edge and $c_b$ is the number of those directed edges.
Upon being a destination node for an edge, the probability response function is defined for a node at a time point $k$, as the stochastic:
\begin{equation}
\label{eq:response}
r_n^{[k]} := \frac{\sum^N_{i=1}l_i (A^{[k]})_{in}}{1+l_N\sum^N_{i=1}(A^{[k]})_{in}}.
\end{equation}
Each time a node will respond, a successful spread of (assumed related content) occurs ($s_n^{[k]}$), creating $c_r$ links uniformly chosen across the whole network. Implied in this response effect is a temporal dependency of content association. It is connected to the motivation for network members to obtain rewards to sharing important information towards other members in a timely fashion.

We add to the model a mechanism of rewarding  participation in stimulating a successful response in another node. This is done by having an increment in node's importance value. By increasing the value $l_n$ this will result in a relatively higher expectation of successful message response in future directed edges produced. We impose that $e_{i,\mathbf{j}}, i \notin \mathbf{j}, |\mathbf{j}| = c_r $ for unique edges to be produced and that they are not self referencing. Any node that has an edge towards another node which produces a  response in $k+1$,  can receive an increase (increment) of their importance value. This occurs regardless of any content presumptions in the model. It focuses solely upon the temporal ordering for the related attributes assumed when incrementing the values. The importance vector is not normalized since the denominator in eq\ref{eq:response}, $1+l_N\sum^N_{i=1}(A^{[k]})_{in}$, provides for a relative measure, and social media platforms in general are not known to scale the analytics of popularity (friend number, retweet numbers etc) according to a network aggregate.

A low ranking node in terms of perceived importance incurs the same increment upon their $l_i$ value when inducing a response through a directed edge. In practice this may differ in how other nodes see incoming messages (content streams) either digitally or through traditional social interactions. We are assuming that from the perspective of the destination node for the content, that the aggregation of all the incoming messages at a particular time point coincides with a similar discussion context. They are then perceived to be relevant in nature for the temporal context (increases in granularity of the samples will reinforce this assumption), and then uniformly attributed for incurring a successful response, $s_n^{[k]}$:
\begin{equation}
\label{eq:increment}
  l_i' = (s_n^{[k]}\times(A^{[k]})_{in} + l_i).
\end{equation}
This provides the motivation for the exchange activity and effort to participate in the influence other nodes in the network. As these increments will enable a user to have longer range of content exposure and recognition feedback.

It can expected that over many iterations this initial max difference in node importance $max(\mathbf{l})-min(\mathbf{l})$ would become less significant. This difference can be assumed to remain constant over multiple simulations and if $max(\mathbf{l})-min(\mathbf{l}) = diff$ is considered to be approximately a constant this value for large $k$ becomes insignificant in comparison to the minimum $min(l_i) >> diff$.
With a uniform addition of importance that is accumulated due to piggy-backing from content and dark retweeting  (removing original authorship labels) the minimization of the initial starting points is anticipated. This reflects the idea of a hypothetical 'ergodic' state destination for a network of users with full visibility over enough time.

The evolution of the importance becomes essential for the study of the results of polarization since there are non-heterogeneous perceptions of content sentiment. As responses induce importance increments to refrain from promoting opposing content, this must not be re-distributed in the network. The nodes having a degree of freedom to reward nodes that more effectively create information specific to a group-think is a mechanism in the dynamics that forms acknowledgement from the group.

To represent a polarization regime; each node is put into one of two different groups, the \emph{odds} and the \emph{evens}. There should be an approximately equal amount of total importance of the nodes for each side and this break in homogeneity affects the response function probabilities by changing the perceived importance from each node. Taking the case where $mod(n,2)=0$, $n$ is on the even labelled group and the polarized importance contributions for its response function becomes:
\begin{equation}
\label{eq:polarImportance}
\mathbf{l}_{even} =
\begin{cases}
  l_i           &  \quad i \text{ is even}\\
  l_i\times(-1) & \quad   i \text{ is odd}
\end{cases}
\end{equation}
and the respective columns in the adjacency matrix used by each calculation of the response function becomes:
\begin{equation}
r_{n\in even}^{[k]} := \frac{\sum^N_{i=1} \left( (A^{[k]})_{*n} \cdot \mathbf{l}_{even}\right)}
{1+l_N\sum^N_{i=1}(A^{[k]})_{in}}
\end{equation}
With the straight forward swapping for the odd case. Having this positive in-group weighting and negative out-group weighting while maintaining the ability for nodes to send messages through transient links is what is explored in the results section. The changes will be manifested in a change in the distribution for the importance values. If the relatively constant value of the difference between the largest and smallest values of the importance changes this would reflect a paradigm shift. Given an understanding of that process a premeditated instantiation could then benefit a member who has more network wealth to lose from a uniform content response probability than the alternative case's results. Given a set of dynamics for the creation and spread of information modeled as messages over discrete time, a stochastic simulation for the spread of information can be produced and the time ordered events analyzed as a temporal network.

\section{Results}
We produce simulated data as described in the methodology section where there are two states of the simulation; a homogeneous community in which content is passed with no differentiation between point of origin of the edges and then a polarized perspective of the importance of content according to  node origin is imposed as in eq~\ref{eq:polarImportance}. The number of nodes chosen is 40 according to the Dunbar number of community sizes where we can expect such a group to be able to message all other members uniformly, \cite{gonccalves2011modeling,dunbar1993coevolution}. This number is also chosen in some of the related research into public discourse on political discussions such as \cite{adamic2005political} which studies 40 blogger accounts that were deemed to be \emph{A-list} in 2004 prior to the US presidential election. The time period for the study was a snapshot of 1K blogs each day and we therefore also use the 1K interval for our model here for the figures (a separate blog analysis of opposing political blogs in \cite{hargittai2008cross} also used 40 blogs which interlinked).

\begin{figure}[h]
  \begin{centering}
\includegraphics[scale=0.525]{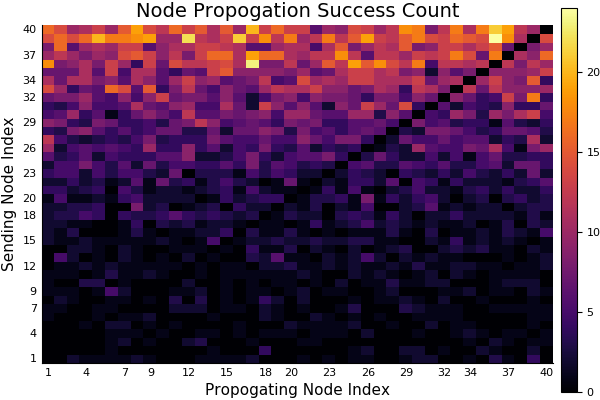}
\includegraphics[scale=0.525]{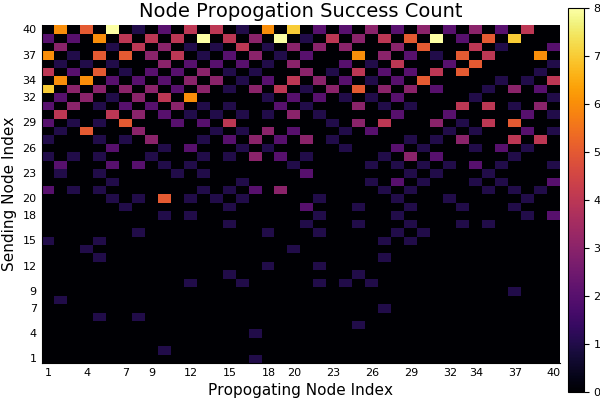}
\caption{\label{fig:heatmaps} Each heatmap shows the simulated accumulation of participation in initiating a response of information dissemination from one node towards a different one $i \rightarrow j$.  On the left is the pre polarization simulation results and on the right post polarization (time points 1-8K, and 8K-16K). We can see that polarization reduces greatly the cross community communication and the total communication produced by nodes lower in the importance ranking. The reduction can be connected to an interpretation of 'intimidation' when messages arrive from opposite sides of the discussion.}
\end{centering}
\end{figure}

Figure~\ref{fig:heatmaps} shows the heatmaps of the before and after polarization simulations to depict the ability of nodes to initiate a content spread response from edges they create or assist in transmitting. The simulation runs for two periods of 8K iterations and shown in each cell is number of responses a node $i$ participated in initiating a response in $j$ by creating an edge towards node $j$ at time $k$ prior to $j$ further creating an edge in $k+1$. This spread of information, by creating links, is counted in each entry and proportional to the color scheme in the legend. The subfigure on the left shows the pre-polarization and the one on the right the post-polarization. The range of values is not rescaled between the two subfigures because in the nature of the communication the relative values within the time periods is the most important feature of the dynamics to investigate. Having a relative difference in communication change is what is assumed as a better indicator of ranked influence than the absolute value. Without introducing a constraint for messages to be exchanged, the polarized state depicts a reduced contribution for spread between the different sides of the network (odds-evens). It also reduces the relative influence of the nodes at the lower end of the spectrum (as is also shown in the next figures) even within their homogeneous polar affiliation.
The heatmap cell score values are calculated using;
\begin{equation}
\label{eq:triggerAvg1}
 score_{i,j} = \sum_{k=1}^{T} s_{i\rightarrow j}^{[k]}.
\end{equation}
Here $s$ is the item indexed array for the successful responses
$r_{i\rightarrow j}^{[k]}$ at time point $k$.

The feature that the nodes which are not ranked on the top of the importance send relatively less messages indicates a phenomenon of 'indimidation'. This can be explained by the numerator of the response function unlikely to produce a firing unless the origin of the content is highly ranked to counter balance opposition in a disagreement. It can loosely be perceived as a defence mechanism to adhere to content spread by the leaders of the fractions when there is doubt originating from another side.

\begin{figure}[h]
  \begin{centering}
  \includegraphics[scale=0.525]{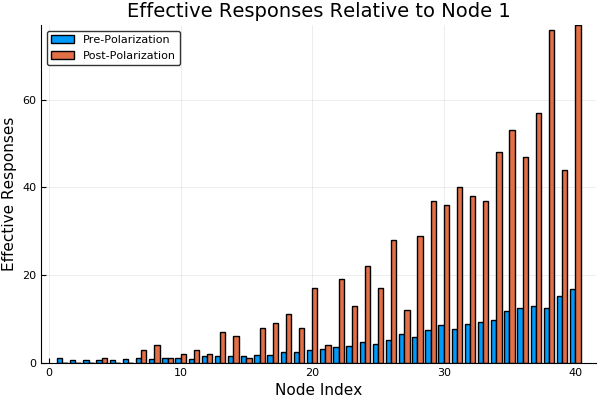}
  \includegraphics[scale=0.525]{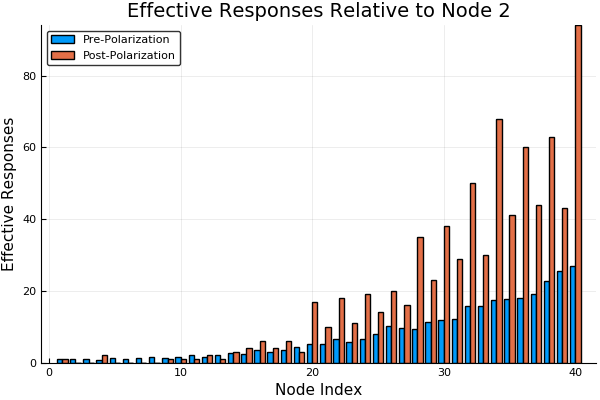}
  \caption{\label{fig:responsesToLast} The plots show the relative total number of effective response events that the nodes have participated in, with relation to the last and second last nodes in terms of initial starting importance. On the left the relative values compared to the last ranked node and on the right to the second last ranked node. The pre and post polarization values are separated to examine the differences between the ends of the message power brokerage spectrum. In blue  are the pre polarization values and in orange the post polarization values. Both plots show the same trend and most importantly that the disparity between the top and bottom nodes increases after polarization.}
  \end{centering}
  \end{figure}

Figure~\ref{fig:responsesToLast} shows investigates the relative values of the network members' ability to trigger an information spread (response). These plots show the response count of all the nodes in relation to the  lowest or second lowest nodes according to how they were initially ranked nodes in the importance values $l_i$. The subplot on the left is for the relative values according to node1 (lowest ranked) and in the subplot on the right is the relative value towards node2 (second lowest ranked). This checks to see if the relationship changes fundamentally across the groups on which a polarization is defined. These relative values are found separately for the polarized case which is in orange and for the homogeneous case which is in blue.
This relative value towards the lowest end gives an idea of the distribution of influence between the top and bottom as a spectrum for the distance of how much more content is spread at the ends of the ranking. The blue bars show that the increase in ranking does result in more response triggers in other nodes, but the polarized case in orange shows a substantially greater increase upon that. The significant feature is that the high ranking nodes increase their ability to send messages towards other nodes in relation to those at the lower end of the spectrum when the network is polarized. This increase in disparity is also seen for the comparison with the lowest member on an opposing member of the discussion. The reason the heights are not completely monotonic across the node numbers is that due to the stochasticity there is room for the nodes to create more or fewer responses in other nodes which incurs reinforcing feedback.
The heights for these bars are measured by the response influence pre and post polarization by the following:
\begin{equation}
\label{eq:triggerAvg1}
 score_{i,T'} = \frac{\sum_{k=1}^{T'} s_{i}^{[k]}}{T'},
\end{equation}
and the ratio of the node triggering capability with another node is:
\begin{equation}
\label{eq:triggerAvg1}
 \frac{score_{i',T'}}{score_{i,T'}} .
\end{equation}
Here we choose $i$ to be values ${1,2}$ and the reason is to show there is not much significant difference between the lower end of the spectrum for each polarized side of the message exchange network.

\begin{figure}[ht]
   \begin{centering}
\includegraphics[scale = 0.65]{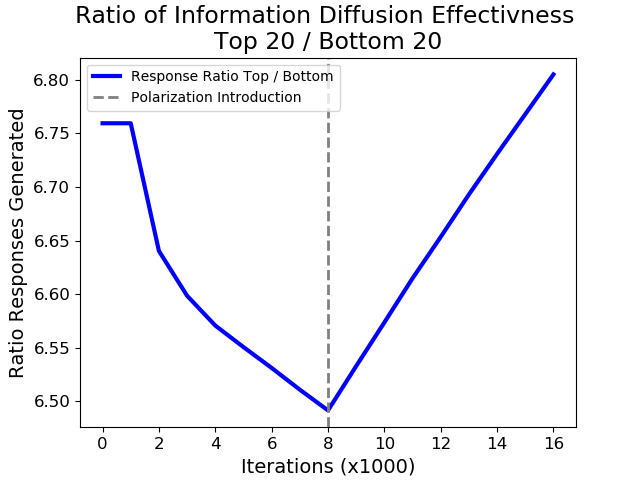}
\caption{\label{fig:tophalfoverbottom}Response initiating power (dissemination power) of the top 20 nodes over the bottom 20 nodes. The first 8000 time points is prior to the polarization of the network and the subsequent time points represents the polarization period. Prior to polari The top 20 nodes gain  increased power once polarization is introduced by using the feature that nodes of less importance cannot cancel out intervening messages from another side of the discussion. Therefore the discussion leaders exclusively manage to increment their importance as a result of the polarization introduction. Such processes will produce benefits for the members which were initially at the top of the important ranking list. (dashed line is where the polarization is introduced)}
 \end{centering}
\end{figure}

Figure~\ref{fig:tophalfoverbottom} shows the effect of polarization on the relative power of the top half of the  influential nodes in respect to the bottom half ${bottom = [1,\ldots,N/2], top = [N/2+1,\ldots,N]}$. The model runs for 8K iterations as a homogeneous network of message exchanges until the subsequent introduction of polarization for the remaining time 8K points. The dashed line identifies the point where the polarization is introduced.
To measure the amount of influence  a certain group of nodes has over another group of nodes to promote a response of message replication, pre and post polarization. To measure the running average of one group's ability to initiate message spread in comparison to other remaining group a sum over the $s_{n}^{[k]}$ values is produced. Given a subset of nodes $N_{top}$ from the total number of nodes $N$ the rest of the nodes in the network are $N_{bottom}=N-N_{top}$. The average number of node message spreading is measured from the first time point to another time point at 1K iteration intervals with the final time point being $T$ and itermediate time points at $T'$  (for the top nodes):
\begin{equation}
\label{eq:triggerAvg1}
 score_{top} = \frac{\sum_{k=1}^{T'}\sum_{n=N_{top}}^{N} s_{n}^{[k]}}{T'}.
\end{equation}
For the ratio of the top to bottom:
\begin{equation}
\label{eq:triggerAvg1}
\frac{score_{top}}{score_{bottom}} = \frac{\sum_{k=1}^{T'}\sum_{n=1}^{N_{top}} s_{n}^{[k]}}{\sum_{k=1}^{T'}\sum_{n=1}^{N_{bottom}} s_{n}^{[k]}}.
\end{equation}
In the simulations here $n_{top}$ is chosen to be $N/2$ and is 20 nodes here.
From the figure it is apparent that the non-polarized regime result in a decreasing trajectory of power disparity. This results in a more even probability distribution for the nodes across the network. In terms of the users it means that all members of the group have a more equal chance of having their content responded to. The interpretation is that polarization introduces  intimidation via uncertainty in the overall accumulation of importance of the content when there are conflicting assignments. The discussion leaders therefore will be of the few able to participate in message responses and then that will increase there importance without the lower half in that time period. This trend post polarizations provides an avenue for the original discussion disparity to be re-instantiated and obtain more control over the spread of content between the users.

\begin{figure}[ht]
   \begin{centering}
\includegraphics[scale = 0.65]{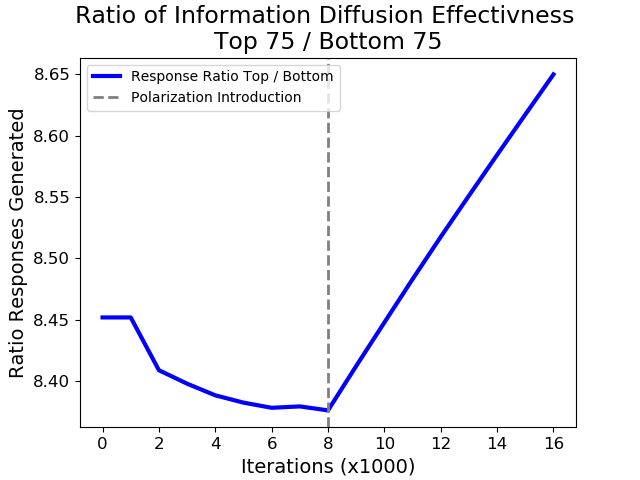}
\caption{\label{fig:tophalfoverbottom150}Response initiating power (dissemination power) of the top 75 nodes over the bottom 75 nodes. The first 8000 time points are prior to the polarization of the network and the following time points represents the polarization period. The top 75 nodes gain  increased response iniaiting power once polarization is introduced by using the feature that nodes of less importance cannot cancel out intervening messages from another side of the discussion. The polarization event benefits the high ranking nodes. (dashed line is where the polarization is introduced)}
 \end{centering}
\end{figure}

In Figure\ref{fig:tophalfoverbottom150} the simulation is performed for the case of 150 nodes which represents the outer reach of the Dunbar circles as it is expected that humans cannot hold more than 150 social ties due to cognitive constraints. This investigation checks the model for the phenomenon observed in Figure\ref{fig:tophalfoverbottom} and whether the top half of the nodes benefit from a polarization event. For this number of users which is the theoretical limit of maintained social ties kept active, we produce the ratio of the top to bottom over the same number of iterations. The polarization point (indicated by the dashed line) is introduced at the same number of iterations, 8K, and the same principal effect is observed. A monotonic decrease of the response ratio of the top vs the bottom  until the polarization event occurs in which the top ranked users in terms of the importance ratio obtain a monotonic increase during the exchange of content where opposition is present. The model therefore shows that the phenomenon that members of the leading importance users for spreading content through other members benefits from polarization and that this is not an artifact associated with the network size.

The ability for a low ranking node to be able to increase its importance influence by participating in  response firings which coincide temporally with high ranking nodes is responsible for their increase in importance over time in a non-polarized discussion. Considering the 40 user case, node 40 and 1 both send a message at the same time with equal uniform probability to a node $i$, and the firing of a sequence of messages is equally attributed to both nodes. It may be seen a fault of the model to not accurately attribute the responsibility for the accumulated effort in the numerator for this activity, but this encapsulates the  feature that nodes can mimic the images of higher ranked nodes with techniques such as 'dark' tweeting (removing credit of origin and repeating \cite{azman2012dark}). Another process is  that  nodes do not separate repeater and origin credit (sharing/retweeting) which is possible for those that simply copy valuable content. This  goes against the Matthew effect of the rich get richer \cite{barabasi1999emergence,perc2014matthew} for the natural occurance in a homogeneous community lacking polarization and complete view. To some degree this reinforces a concept introduced in \cite{katz1955lazarsfeld} that frequent interaction leads to attitude conformity.

From the point where polarization is introduced we can see that the accumulated power of the top half of the network nodes begins to consistently increase over their historical average. The reasoning is that given an expected number of messages received from other nodes in the network, there will be an equal likelihood of nodes in the same group and opposing group, and that only in the situations where the nodes participating from the same group have a large importance will the overall contribution to fire be positive and produce a response. This models a type of competition for activating nodes to trigger a firing and only happens when there is a large positive perception of the current status of the incoming messages. This incoming summation of importance is one-sided sentiment which perceives certain nodes as positive and the rest as negative. This coincides with the observations of \cite{adamic2005political} which saw that there is clustering between the groups and differences in the interlinking due to the edge dynamics. The edge dynamics there produced in the analyzed data a similar effect through a simple competitive importance reward iteration. The work of \cite{hargittai2008cross} has a similar static image of the polarized states during the following political cycle of 2008, where the figures show a limited number of interlinking between communities and indications of focal points of edges within each of the two groups. The continual growth of the influential nodes where those on the bottom half are continuously reducing their response rate makes them more 'passive' as they receive but not spread content in an influential manner. This reflects the findings in \cite{romero2011influence} from twitter datasets which differentiates influential from popular; as the response mechanism here measures a similar tweet mechanism where influence includes a mechanism of spreading.

We can see that polarization occurs within an environment where there is room for disagreement and perceived importance plays a role in overcomming a threshold for relaying the message to the rest of the network. In effect we are simulating through a modification of a general stochastic framework the perception of \emph{confidence}. This confidence factor is in its basic form here, the overall importance of the current content received at a given time point. There can be identifiability considerations; such as the variability of the reference to the polarized discussion topics which are interleaved, but these are omitted as a possible extension.

%There is a  chance that a basal rate fire (content creation step) coincides with an external content dissemination step, but as the compound probability of this is low and it is difficult to filter out these false positives from data collection without dissecting the content which is also not always accurate the occurance is not accounted for. Here the $s$ records the success of a 'response' towards disseminating content towards other nodes.

\section{Conclusions}
The framework of Dynamic Communicators \cite{mantzaris2012model} provides a basic general framework to explore the ability of nodes to transit content to each other in a stochastic message passing network. The extent of how a node can 'influence' another node it is connected to is measured by the number of responses it can initiate. This influence measurement is represented by a parameter of importance which is allowed to increase with a successful edge initiation response from the transient edges produced. The analysis of the simulations  examined changes in the power brokerage between nodes when a polarization event is introduced.
The simulations measure the disparity between the ability for nodes to have their content spread by intermediates. The initial relative values of the largest importance value holders, in comparison with the lower end of the spectrum  are shown over iterations of the stochastic dynamics. This process of representing importance reflects many scenarios such as  increasing the responsiveness of nodes towards those with close proximity to information of rich content that certain members initially bring from an external environment. Other reasons could exist such as a familiarity with all users in networks where a complete view is possible (this relates to networks without imposed hierarchical structure).

In summary the results show that over time, members of the network, through incremental improvements in their perceived importance, develop a more uniform rate of content responses from their peers. When a polarization event is introduced this incurs a penalty for the lower ranked nodes to be able to have their content disseminated and the disparity between the high ranked and low ranked users increases. This mirrors our intuitive understanding that networks without shocks should stabilize to a more uniform distribution over time.
 Future work would entail examining more intricate network connectivities that reflect the environmental costs for various nodes to communicate together in. Reflecting upon the well known case study of the Zachary Karate club \cite{zachary1977information}, this provides insight into the plausible insentives the leaders of the two fractions of the disagreement may have had regardless of the self reported motives.

\bibliographystyle{apalike}
\bibliography{polarRefs.bib}

\end{document}